\documentclass[twocolumn,showpacs,amsfonts,aps,prc,floatfix,nofootinbib,float,superscriptaddress]{revtex4}
\usepackage{amsmath}
\usepackage{bm}
\usepackage{graphicx}
\usepackage{color}
\usepackage{booktabs}

\usepackage{placeins}
\usepackage{enumerate}
\usepackage{MnSymbol}

\begin{document}
\title{Langevin dynamics and decoherence of heavy quarks at high temperatures}
\date{\today}

\author{Yukinao Akamatsu}
\affiliation{Kobayashi-Maskawa Institute for the Origin of Particles and the Universe (KMI), Nagoya University, Nagoya 464-8602, Japan}

\begin{abstract}
A Langevin equation of heavy quarks in high-temperature quark-gluon plasma is derived.
The dynamics of heavy quark color is coupled with the phase space dynamics and causes a macroscopic superposition state of heavy quark momentum.
Decoherence of the superposition state allows one to use a classical description.
The time scale of decoherence gives an appropriate discretization time scale $\Delta t\sim \sqrt{M/C_{\rm F}\gamma}$ for the classical Langevin equation, where $M$ is heavy quark mass and $\gamma$ is heavy quark momentum diffusion constant.
\end{abstract}


\maketitle

\section{Introduction}
The transport properties of quark-gluon plasma (QGP) have attracted a lot of interest since nearly perfect liquid behavior has been discovered in the relativistic heavy-ion collision experiments at the Relativistic Heavy Ion Collider (RHIC) \cite{Teaney:2009qa}.
Further investigations in the heavy-ion collisions at higher energy are ongoing at the Large Hadron Collider (LHC).
The discovery of nearly perfect liquid behavior seems to contradict the notion of weakly interacting QGP, and it rather suggests a strongly interacting nature of the QGP.
Indeed, the universal lower bound of the ratio of shear viscosity to entropy density $\eta/s\geq 1/4\pi$ is proposed in the strongly interacting quantum field theories based on the conjectured duality between the gauge theory and string theory, or the anti-de-Sitter space and conformal field theory (AdS/CFT) correspondence \cite{Kovtun:2004de}.
This is close to the value $\eta/s\sim (1\-3)/4\pi$, which is phenomenologically extracted from the experimental data by hydrodynamic simulations for the heavy-ion collisions \cite{Song:2012ua}.

In the heavy-ion collisions, not only the bulk collective dynamics but also hard probes, such as heavy quarks and jets, reveal independent aspects of the transport properties of the QGP.
For example, medium modification of heavy quark momentum spectra offers an opportunity to study drag force of the QGP acting on a test particle with color \cite{Moore:2004tg}.
Phenomenological studies of the heavy quark probe in the heavy-ion collisions are summarized in Ref.~\cite{Rapp:2009my}.
Since there is a kinematical hierarchy between the heavy quark and the QGP, the heavy quark dynamics is slow compared to the correlation time of matter constituents so that the effects of interaction between them are averaged out.
This enables one to use a simple effective description of the heavy quark using the drag and fluctuation forces.
There have been various efforts to calculate the heavy quark transport coefficients by the perturbation theory \cite{Moore:2004tg,CaronHuot:2007gq}, by the lattice QCD simulations \cite{Petreczky:2005nh}, and by applying the AdS/CFT correspondence \cite{Herzog:2006gh}.

In this paper, I shed light on the dynamics in the heavy quark color space.
The Langevin dynamics is a classical description of the Brownian motion.
Unlike kinetic variables, the time scale of the heavy quark color does not get long even with its heavy mass.
Typically, the time scale of color diffusion is about $1/g^2T$ while the duration of a soft scattering is about $1/gT$.
Therefore, it is only in the weak coupling regime that one can expect the validity of Langevin description that couples with the heavy quark color degrees of freedom.
In the case of the heavy quark with color, description in the phase space is classical while that in the color space is quantum.
Here the heavy quark color is in the fundamental representation of color ${\rm SU}(N_{\rm c})$.
Instead of describing it classically by Wong's equation \cite{Wong:1970fu} in terms of a $(N_{\rm c}^2-1)$-dimensional vector, I treat a color state as a quantum state in a Hilbert space with $N_{\rm c}$ dimensions.
Indeed, the latter description is a direct consequence of the quantum description of the heavy quark Brownian motion \cite{Akamatsu:2014qsa}.

The main finding of this paper is that an analog of the Schr\"odinger's cat state appears in the heavy quark Langevin dynamics due to the non-Abelian interaction of QCD.
The classical momentum corresponds to the cat and couples to quantum states in the color space.
It is well known that decoherence of such a macroscopic superposition state is essential to have a classical picture of the Brownian motion \cite{Caldera:1985tk}.
Consequently, the discretization time scale of the classical Langevin dynamics must come out from the time scale of decoherence in the heavy quark sector.
The discretization time scale turns out to be $\Delta t\sim\sqrt{M/C_{\rm F}\gamma}$.
It depends on the heavy quark mass $M$ and on the heavy quark momentum diffusion constant $\gamma$ (up to some factors).

Phenomenological implication of the discretization time scale $\Delta t\sim\sqrt{M/C_{\rm F}\gamma}$ is intriguing.
Typically, the drag force extracted from the experimental data corresponds to $C_{\rm F}\gamma\sim 1 T^3$ \cite{Rapp:2009my}.
For a charm (bottom) quark in the QGP with $T\sim 200 \ {\rm MeV}$, the discretization time scale is $\Delta t\sim 3 \ {\rm fm} \ (5 \ {\rm fm})$.
In the heavy-ion collisions, typical lifetime of the QGP is $\tau_{\rm QGP}\sim 10 \ {\rm fm}$.
Therefore the values of $\Delta t$ might indicate that the macroscopic superposition state remains, at least partially, until the freeze-out stage of the heavy-ion collisions.
Also the heavy quark hadronization time scale $1/\Lambda_{\rm QCD}\sim 1 \ {\rm fm}$ is shorter than $\Delta t$ and thus in any case the freezeout process may be able to resolve the macroscopic superposition state.
This would enhance the possibility of recombining heavy quark$-$antiquark pairs into heavy quarkonia and might support the statistical hadronization models \cite{BraunMunzinger:2000px} when heavy quarks are produced in abundance.

This paper is organized as follows.
In Sec.~\ref{sec:HQ_open_system}, I introduce and analyze the basic properties of the heavy quark master equation.
One can see that Ehrenfest equations show kinetic equilibration of heavy quarks.
In Sec.~\ref{sec:derivation_Langevin}, I derive the Langevin equation of heavy quarks with color.
The Langevin equation is not closed in the heavy quark phase space: The momentum update depends on the heavy quark color state.
In Sec.~\ref{sec:Langevin_decoherence}, I discuss how to interpret the coupling between the dynamics in the heavy quark phase space and that in the color space.
Also the appropriate discretization time scale for classical Langevin description is discussed.
Section \ref{sec:summary} is devoted to a summary.
Throughout this paper, I adopt the natural units, $\hbar=c=k_{\rm B}=1$, and operators in Hilbert and Fock spaces are denoted by bold fonts.

\section{Heavy quarks as an open quantum system}
\label{sec:HQ_open_system}
When one describes a heavy quark in the quark-gluon plasma (QGP) as an open quantum system \cite{BrePetText}, reduced density matrix
\begin{eqnarray}
\bm\rho_{\rm Q}(t)={\rm Tr}_{\rm med} \bm\rho_{\rm tot}(t)
\end{eqnarray}
is a basic dynamical quantity.
Here $\bm\rho_{\rm tot}(t)$ denotes a density matrix of a total system of the heavy quark and the QGP.
The total Hilbert space is composed of a direct product of the heavy quark and the QGP Hilbert spaces ${\mathcal H}_{\rm tot} = {\mathcal H}_{\rm Q}\otimes {\mathcal H}_{\rm med}$.
By taking the trace over the Hilbert space for the QGP (${\rm Tr}_{\rm med}$), one gets the reduced density matrix $\bm \rho_{\rm Q}(t)$ that operates in the heavy quark Hilbert space.

The master equation for the reduced density matrix $\bm\rho_{\rm Q}(t)$ was derived by the influence functional formalism \cite{Akamatsu:2014qsa}.
In the leading order of the QCD coupling $g$ and in the nonrelativistic limit,
\footnote{
In this paper, I need not start from the Lindblad form of the master equation which preserves the positivity of the reduced density matrix.
Therefore I neglect all the terms with $A(\vec r)$ in Ref.~\cite{Akamatsu:2014qsa}.
}
it is
\begin{eqnarray}
\label{eq:master}
&&\frac{\partial}{\partial t}\hat\rho_{\rm Q}(t,\vec x,\vec y)
=i\frac{\vec\nabla_x^2-\vec\nabla_y^2}{2M}\hat\rho_{\rm Q}(t,\vec x,\vec y)\nonumber\\
&& \ \ \ + \ F_1(\vec x-\vec y)\hat t^a\hat\rho_{\rm Q}(t,\vec x,\vec y)\hat t^a
-C_{\rm F}F_1(\vec 0)\hat\rho_{\rm Q}(t,\vec x,\vec y)\nonumber \\
&& \ \ \ + \ \vec F_2(\vec x-\vec y)\cdot (\vec \nabla_x -\vec \nabla_y)\hat t^a\hat\rho_{\rm Q}(t,\vec x,\vec y)\hat t^a.
\end{eqnarray}
Here $[\hat\rho_{\rm Q}(t,\vec x,\vec y)]_{ij} = \langle \vec x,i|\bm\rho_{\rm Q}(t)| \vec y,j\rangle$ is a reduced density matrix in the position and color spaces.
The matrix $\hat t^a$ is in the fundamental representation of the color ${\rm SU}(N_{\rm c})$ algebra and $C_{\rm F}=(N_{\rm c}^2-1)/2N_{\rm c}$.
$F_1(\vec r)$ and $\vec F_2(\vec r)$ are given in terms of a function $D(\vec r)$ as
\begin{eqnarray}
F_1(\vec r)&=& -\left(D(\vec r)+\frac{\vec\nabla^2 D(\vec r)}{4MT}\right),\\
\vec F_2(\vec r)&=& -\vec\nabla\left(\frac{D(\vec r)}{4MT}\right).
\end{eqnarray}
Here $D(\vec r)$ is defined by a thermal two-point function of gluons
\begin{eqnarray}
D(\vec r)&\equiv & -\frac{g^2}{N_{\rm c}^2-1}\int_{-\infty}^{\infty} dt \left\langle A^a_0(t,\vec r)A^a_0(0,\vec 0)  \right\rangle_T.
\end{eqnarray}
As is clear from the definition, $D(\vec r)$ is an even function of $\vec r$.
At $|\vec r|\sim 1/gT$, the hard thermal loop resummed perturbation theory gives at leading order
\begin{eqnarray}
D(\vec r)&=&-g^2T\int\frac{d^3k}{(2\pi)^3}
\frac{\pi\omega_{\rm D}^2e^{i\vec k\cdot\vec r}}{k(k^2+\omega_{\rm D}^2)^2}.
\end{eqnarray}
Here $\omega_{\rm D}$ is the Debye screening mass $\omega_{\rm D}^2=(g^2T^2/3)(N_{\rm c}+N_{\rm f}/2)$ for QCD with $N_{\rm f}$ light flavors.

From the master equation \eqref{eq:master}, one can derive time-evolution of thermal averaged quantum expectation values $\llangle \bm{\mathcal O}\rrangle(t)={\rm Tr_{\rm Q}}\left\{\bm\rho_{\rm Q}(t) \bm{\mathcal O}\right\}$, or the Ehrenfest equations, such as
\begin{eqnarray}
\label{eq:xdot}
\frac{d}{dt} \llangle\vec {\bm x}\rrangle
&=&\frac{\llangle \vec {\bm p}\rrangle}{M}, \\
\label{eq:pdot}
\frac{d}{dt} \llangle \vec{\bm p} \rrangle
&=&-\frac{C_{\rm F}\gamma}{2MT} \llangle \vec{\bm p} \rrangle, \\
\label{eq:Edot}
\frac{d}{dt} \left\llangle \frac{\vec{\bm p}^2}{2M} \right\rrangle
&=&-\frac{C_{\rm F}\gamma}{MT}\left(\left\llangle \frac{\vec{\bm p}^2}{2M} \right\rrangle
-\frac{3T}{2}(1+\epsilon)\right),
\end{eqnarray}
reproducing the {\it consequences} of classical Langevin dynamics (up to $\epsilon$).
The drag force parameter $\gamma>0$ and the correction to equipartition $\epsilon$ $(|\epsilon|\ll1)$ are given by
\begin{eqnarray}
\gamma &=& \frac{1}{3}\vec\nabla^2 D(\vec r)|_{r\to 0},\\
\epsilon &=& \frac{1}{4MT}\frac{(\vec\nabla^2)^2 D(\vec r)|_{r\to 0}}{\vec\nabla^2 D(\vec r)|_{r\to 0}}.
\end{eqnarray}

The master equation \eqref{eq:master} has several important properties, which I summarize here.
The terms with $F_1$ describe thermal fluctuation and that with $\vec F_2$ describes dissipation.
In the Ehrenfest equations \eqref{eq:pdot} and \eqref{eq:Edot}, the thermal fluctuation $\frac{C_{\rm F}\gamma}{MT}\frac{3T}{2}(1+\epsilon)$ derives from $F_1$ whereas the damping terms $-\frac{C_{\rm F}\gamma}{2MT}\llangle\vec{\bm p}\rrangle$ and $-\frac{C_{\rm F}\gamma}{MT}\left\llangle\frac{\vec{\bm p}^2}{2M}\right\rrangle$ are from $\vec F_2$.
By taking the trace in the color space, the color averaged master equation for $\bar\rho_{\rm Q}(t,\vec x,\vec y)\equiv {\rm tr}\{\hat\rho_{\rm Q}(t,\vec x,\vec y)\}$ is derived as
\begin{eqnarray}
\label{eq:master_ctraced}
\frac{\partial}{\partial t}\bar\rho_{\rm Q}(t,\vec x,\vec y)
&=&i\frac{\vec\nabla_x^2-\vec\nabla_y^2}{2M}\bar\rho_{\rm Q}(t,\vec x,\vec y)\\
&& + \ C_{\rm F}\left(F_1(\vec x-\vec y)-F_1(\vec 0)\right)\bar\rho_{\rm Q}(t,\vec x,\vec y) \nonumber \\
&& + \ C_{\rm F}\vec F_2(\vec x-\vec y)\cdot (\vec \nabla_x -\vec \nabla_y)\bar\rho_{\rm Q}(t,\vec x,\vec y).\nonumber
\end{eqnarray}
Close to equilibrium, thermal de Broglie wavelength for a heavy quark $\l_{\rm dB}\sim 1/\sqrt{MT}$ is much shorter than typical correlation length of (electric) gluons $\l_{\rm fluct}\sim 1/gT$, where for $|\vec r|\agt l_{\rm fluct}$, $D(\vec r)\simeq 0$.
Therefore one can take a short distance approximation for $D(\vec r)\simeq D_0 + (D_2/6) \vec r^2$.
In this approximation, $\epsilon=0$.
By this expansion, one gets the Caldeira-Leggett master equation \cite{Caldeira:1982iu}.

In the following, I derive the Langevin equation itself, instead of being satisfied with reproducing the $\it consequences$ of its dynamics.

\section{Derivation of Langevin dynamics}
\label{sec:derivation_Langevin}
\subsection{Stochastic master equation}
First, let us investigate more closely how the $F_1$ terms in the master equation \eqref{eq:master} are related to the thermal fluctuation.
Suppose an infinitesimal time step $t\to t+dt$ under a random background $\xi^a(t,\vec x)$, which rotates a heavy quark color state by $\hat U_{\xi}(t,\vec x)\equiv\exp[-idt\xi^a(t,\vec x) \hat t^a]$.
The following statistical property of the random field is assumed:
\begin{eqnarray}
\label{eq:xi_statistics}
\left\{
\begin{aligned}
\langle \xi^a(t,\vec x)\xi^b(s,\vec y) \rangle_T
&=F_1(\vec x-\vec y)\delta(t-s)\delta^{ab}, \\
\langle \xi^a(t,\vec x)\rangle_T&=0.
\end{aligned}
\right.
\end{eqnarray}
Hereafter, I call average over the random field $\xi$ the noise average and denote it as $\langle\cdots\rangle_{\xi}$.
Then the density matrix in the random field, which I denote as $\hat\rho_{\rm Q}(t,\vec x,\vec y;\xi)$, evolves as
\begin{eqnarray}
\label{eq:master_xi}
&&\hat\rho_{\rm Q}(t+dt,\vec x,\vec y;\xi)\nonumber\\
&&=\hat U_{\xi}(t,\vec x)
\hat\rho_{\rm Q}(t,\vec x,\vec y;\xi)
\left[\hat U_{\xi}(t,\vec y)\right]^{\dagger}.
\end{eqnarray}
Noting that $\xi\propto 1/dt^{1/2}$, $\hat U_{\xi}(t,\vec x)$ can be expanded as
\begin{eqnarray}
\label{eq:U_xi}
\hat U_{\xi}(t,\vec x)
&=&1-idt\xi^a(t,\vec x) \hat t^a
-\frac{dt^2}{2}\left[\xi^a(t,\vec x) \hat t^a\right]^2 + \mathcal O(dt^{3/2}) \nonumber \\
&\simeq&1-idt\xi^a(t,\vec x) \hat t^a-\frac{dt}{2}C_{\rm F}F_1(\vec 0).
\end{eqnarray}
In the second line of Eq.~\eqref{eq:U_xi}, I approximate $dt^2\left[\xi^a(t,\vec x) \hat t^a\right]^2\simeq dt^2\langle\left[\xi^a(t,\vec x) \hat t^a\right]^2\rangle_{\xi}=dtC_{\rm F}F_1(\vec 0)$ because the fluctuation around the average is $\propto dt$.
Such a fluctuation can be neglected, for it does not contribute to the master equation after taking the noise average.
Therefore one can always substitute $dt^2\xi\xi\simeq dt^2\langle\xi\xi\rangle_{\xi}$.
The evolution equation \eqref{eq:master_xi} then becomes
\begin{eqnarray}
\label{eq:master_xi2}
&&\frac{\partial}{\partial t} \hat\rho_{\rm Q}(t,\vec x,\vec y;\xi)\\
&& = F_1(\vec x-\vec y)\hat t^a\hat\rho_{\rm Q}(t,\vec x,\vec y;\xi)\hat t^a
 - \ C_{\rm F}F_1(\vec 0)\hat\rho_{\rm Q}(t,\vec x,\vec y;\xi)\nonumber\\
&& \ \ \ - \ i\left[
\xi^a(t,\vec x) \hat t^a\hat\rho_{\rm Q}(t,\vec x,\vec y;\xi)
-\hat\rho_{\rm Q}(t,\vec x,\vec y;\xi)\xi^a(t,\vec y)\hat t^a
\right].
\nonumber
\end{eqnarray}
Taking the noise average for Eq.~\eqref{eq:master_xi2} and interpreting the noise-averaged density matrix $\langle\hat\rho_{\rm Q}(t,\vec x,\vec y;\xi)\rangle_{\xi}$ as the original density matrix $\hat\rho_{\rm Q}(t,\vec x,\vec y)$, one can reproduce the thermal fluctuation terms ($F_1$ terms) in the master equation \eqref{eq:master}.
Adding the kinetic term and the dissipation terms ($\vec F_2$ terms) to Eq.~\eqref{eq:master_xi2}, I derive the following stochastic master equation:
\begin{eqnarray}
\label{eq:stochastic_master}
&&\frac{\partial}{\partial t}\hat\rho_{\rm Q}(t,\vec x,\vec y;\xi)
=i\frac{\vec\nabla_x^2-\vec\nabla_y^2}{2M}\hat\rho_{\rm Q}(t,\vec x,\vec y;\xi)\\
&& \ \ \ + \ F_1(\vec x-\vec y)\hat t^a\hat\rho_{\rm Q}(t,\vec x,\vec y;\xi)\hat t^a
-C_{\rm F}F_1(\vec 0)\hat\rho_{\rm Q}(t,\vec x,\vec y;\xi)\nonumber \\
&& \ \ \ + \ \vec F_2(\vec x-\vec y)\cdot (\vec \nabla_x -\vec \nabla_y)\hat t^a\hat\rho_{\rm Q}(t,\vec x,\vec y;\xi)\hat t^a\nonumber\\
&& \ \ \ - \ i\left[
\xi^a(t,\vec x) \hat t^a\hat\rho_{\rm Q}(t,\vec x,\vec y;\xi)
-\hat\rho_{\rm Q}(t,\vec x,\vec y;\xi)\xi^a(t,\vec y)\hat t^a
\right].\nonumber
\end{eqnarray}

Although I construct the stochastic master equation \eqref{eq:stochastic_master} quite intuitively, there also exists a formal derivation.
One can formulate the problem using the influence functional formalism \cite{Feynman:1963fq}.
Following the notation of Ref.~\cite{Akamatsu:2014qsa}, the influence functional $S_{\rm IF}$ can be obtained as an expansion $S_{\rm IF}=S_{\rm pot}+S_{\rm fluct}+S_{\rm diss}+\cdots$.
The thermal fluctuation $S_{\rm fluct}$ and a part of the dissipation $S_{\rm diss}$ can be expressed using a stochastic variable equivalent to $\xi^a$ here.
See Appendix \ref{sec:thermal_fluctuation} for details.

\subsection{Langevin equation with color}
Using the stochastic master equation \eqref{eq:stochastic_master}, one can derive stochastic Ehrenfest equations.
Since ${\rm Tr_{\rm Q}}\left\{\bm\rho_{\rm Q}(t;\xi) \bm{\mathcal O}\right\}$ depends on the random field $\xi^a$, it corresponds to quantum expectation value in a particular realization of $\xi^a$, which I denote as $\langle\bm{\mathcal O}\rangle(t;\xi)$ or simply as $\langle\bm{\mathcal O}\rangle(t)$ if there is no confusion.
The relation to the thermal average of quantum expectation value is $\llangle\bm{\mathcal O}\rrangle(t)=\langle\left[\langle\bm{\mathcal O}\rangle(t;\xi)\right]\rangle_{\xi}$.
The following stochastic Ehrenfest equations are derived:
\begin{eqnarray}
\label{eq:xdot_xi}
\frac{d}{dt} \langle\vec {\bm x}\rangle
&=&\frac{\langle \vec {\bm p}\rangle}{M}, \\
\label{eq:pdot_xi}
\frac{d}{dt} \langle \vec{\bm p} \rangle
&=&-\frac{C_{\rm F}\gamma}{2MT} \langle \vec{\bm p} \rangle
+ \vec f^a(t)\langle \bm{t^a}\rangle, 
\end{eqnarray}
with
\begin{eqnarray}
\label{eq:kick_strength}
\langle f^a_i(t) f^b_j(t')\rangle_T=\gamma(1+\epsilon)\delta(t-t')\delta^{ab}\delta_{ij}.
\end{eqnarray}

In the derivation, I assume that the heavy quark is localized as a wave packet.
To be explicit, the spatial extension of the wave packet should be much smaller than the correlation length of gluons $l_{\rm fluct}$.
This is the classical (point particle) limit for the heavy quark.
In such a limit, it is natural to assume that the density matrix is factorized in the color and the configuration spaces ${\bm \rho}_{\rm Q}(t;\xi)\approx {\bm\rho}_{\rm color}(t;\xi)\otimes{\bm \rho}_{\rm conf}(t;\xi)$.
\footnote{
Although I denote the color space density matrix by an operator in the Hilbert space ${\bm \rho}_{\rm color}$, it is defined only in the $N_{\rm c}\times N_{\rm c}^*$ representation.
}
In terms of ${\bm\rho}_{\rm color}$, $\langle {\bm t^a}\rangle(t)={\rm Tr}_{\rm color}\left\{{\bm\rho}_{\rm color}(t;\xi){\bm t^a}\right\}$.
The random force $\vec f^a$ is the force in the random potential $\xi^a$ evaluated at the position of wave packet $\vec x=\langle \vec {\bm x}\rangle(t)$:
\begin{eqnarray}
\label{eq:kick}
\vec f^a(t)\equiv -\vec\nabla \xi^a(t,\vec x=\langle \vec {\bm x}\rangle(t)).
\end{eqnarray}
The random force strength \eqref{eq:kick_strength} follows from the definitions \eqref{eq:xi_statistics} and \eqref{eq:kick}.

Unlike the conventional Langevin equation, the ones derived here in Eqs.~\eqref{eq:xdot_xi} and \eqref{eq:pdot_xi} are not in a closed form.
One needs to know the dynamics of $\langle {\bm t^a}\rangle$.
Using the same assumption as made for ${\bm \rho}_{\rm Q}(t;\xi)$, it is derived as
\begin{eqnarray}
\label{eq:tadot_xi}
\frac{d}{dt}\langle{\bm t^a}\rangle&=&
-\frac{\alpha}{2} f^{abc}f^{bcd}\langle{\bm t^d}\rangle
+f^{abc}\zeta^b(t)\langle{\bm t^c}\rangle,
\end{eqnarray}
with
\begin{eqnarray}
\langle\zeta^a(t)\zeta^b(t')\rangle_T&=&\alpha\delta(t-t')\delta^{ab}.
\end{eqnarray}
Here $\zeta^a$ is the random field evaluated at the position of the wave packet $\zeta^a(t)\equiv\xi^a(t,\vec x=\langle\vec{\bm x}\rangle(t))$.
Note also that $\langle f^a_i(t)\zeta^b(t')\rangle_T=0$ which follows from the definitions of $\vec f^a$ and $\zeta^a$.
The parameter is $\alpha=F_1(\vec 0)>0$ and $f^{abc}$ is structure constant of the color ${\rm SU}(N_{\rm c})$ algebra.
Hereafter, I write ${\bm \rho}_{\rm color}(t;\zeta)$ instead of ${\bm \rho}_{\rm color}(t;\xi)$.

The stochastic equation \eqref{eq:tadot_xi} has an interesting property: It conserves $\langle{\bm t^a}\rangle\langle{\bm t^a}\rangle$.
If and only if ${\bm\rho}_{\rm color}(t;\zeta)$ is a pure state density matrix, $\langle{\bm t^a}\rangle\langle{\bm t^a}\rangle=\frac{1}{2}\left(1-\frac{1}{N_{\rm c}}\right)$.
Therefore ${\bm\rho}_{\rm color}(t;\zeta)$ stays a pure state density matrix if it was initially.
This feature can also be understood concisely by reconstructing $\left[\hat\rho_{\rm color}(t;\zeta)\right]_{ij}\equiv\langle i|{\bm\rho}_{\rm color}(t;\zeta)|j\rangle$ with
\begin{eqnarray}
\hat\rho_{\rm color}(t;\zeta)=\frac{1}{N_{\rm c}}+\left[2\langle{\bm t^a}\rangle(t)\right] \hat t^a.
\end{eqnarray}
Equation \eqref{eq:tadot_xi} is equivalent to
\begin{eqnarray}
\label{eq:colormaster_xi}
\frac{d}{dt}\hat\rho_{\rm color}(t;\zeta)
&=&\alpha \left[
\hat t^a\hat\rho_{\rm color}(t;\zeta)\hat t^a-C_{\rm F}\hat\rho_{\rm color}(t;\zeta)
\right]\nonumber\\
&& - \ i\left[
\zeta^a(t)\hat t^a,\hat\rho_{\rm color}(t;\zeta)
\right].
\end{eqnarray}
This color space master equation is very similar to Eq.~\eqref{eq:master_xi2} with substitution $\vec x=\vec y=\langle \vec{\bm x}\rangle(t)$.
Therefore Eq.~\eqref{eq:colormaster_xi} can be derived from random rotation in the color space $\hat\rho_{\rm color}(t+dt;\zeta)=\hat U_{\zeta}(t)\hat\rho_{\rm color}(t;\zeta)\left[\hat U_{\zeta}(t)\right]^{\dagger}$ with $\hat U_{\zeta}(t)\equiv\exp[-idt\zeta^a(t)\hat t^a]$.
Since the random rotation $\hat U_{\zeta}$ is a unitary evolution, one obtains ${\rm Tr}_{\rm color}\left\{\left[{\bm \rho}_{\rm color}(t+dt;\zeta)\right]^2\right\}={\rm Tr}_{\rm color}\left\{\left[{\bm \rho}_{\rm color}(t;\zeta)\right]^2\right\}$, which gives the conservation of $\langle{\bm t^a}\rangle\langle{\bm t^a}\rangle$.
Note that ${\rm Tr}_{\rm color}\left\{\left[{\bm \rho}_{\rm color}(t;\zeta)\right]^2\right\}\leq {\rm Tr}_{\rm color}\left\{{\bm \rho}_{\rm color}(t;\zeta)\right\}=1$ sets the upper limit of $\langle{\bm t^a}\rangle\langle{\bm t^a}\rangle$.
Because of the conservation of ${\rm Tr}_{\rm color}\left\{\left[{\bm \rho}_{\rm color}(t;\zeta)\right]^2\right\}$, it is natural to demand (only initially) that ${\bm \rho}_{\rm color}(t;\zeta)$ be a pure state density matrix.

Let us summarize what I have obtained so far.
Taking the classical point particle limit for the heavy quark, I have derived the coupled Langevin equations in the phase space and in the color space:
\begin{eqnarray}
\label{eq:color_Langevin}
\left\{
\begin{aligned}
\frac{d}{dt} \langle\vec {\bm x}\rangle
&=\frac{\langle \vec {\bm p}\rangle}{M},\\
\frac{d}{dt} \langle \vec{\bm p} \rangle
&=-\frac{C_{\rm F}\gamma}{2MT} \langle \vec{\bm p} \rangle
+ \vec f^a(t)\langle \bm{t^a}\rangle, \\
\frac{d}{dt}\langle{\bm t^a}\rangle
&=-\frac{\alpha}{2} f^{abc}f^{bcd}\langle{\bm t^d}\rangle
+f^{abc}\zeta^b(t)\langle{\bm t^c}\rangle.
\end{aligned}
\right.
\end{eqnarray}
The statistical properties of the noises are
\begin{eqnarray}
\left\{
\begin{aligned}
\langle f^a_i(t) f^b_j(t')\rangle_T&=\gamma(1+\epsilon)\delta(t-t')\delta^{ab}\delta_{ij},\\
\langle\zeta^a(t)\zeta^b(t')\rangle_T&=\alpha\delta(t-t')\delta^{ab},\\
\langle f^a_i(t)\zeta^b(t')\rangle_T&=0. 
\end{aligned}
\right.
\end{eqnarray}
The initial condition for $\langle {\bm t^a}\rangle$ should satisfy the pure state constraint $\langle {\bm t^a}\rangle\langle{\bm t^a}\rangle=\frac{1}{2}\left(1-\frac{1}{N_{\rm c}}\right)$.
In the above equations, the noises $\vec f^a$ and $\zeta^a$ derive from $\xi^a$.
Therefore I still use the notation of $\langle\cdots\rangle_{\xi}$ for taking the noise average over $\vec f^a$ and $\zeta^a$.

Typical magnitudes of the parameters are
\begin{eqnarray}
\left\{
\begin{aligned}
\gamma&=\frac{1}{3}\vec\nabla^2 D(\vec r)|_{r\to 0}
\sim g^4\ln(1/g)T^3, \\
\epsilon&=\frac{1}{4MT}\frac{(\vec\nabla^2)^2 D(\vec r)|_{r\to 0}}{\vec\nabla^2 D(\vec r)|_{r\to 0}}\propto \frac{T}{M} \ll 1,\\
\alpha&=F_1(\vec 0)\sim g^2T.
\end{aligned}
\right.
\end{eqnarray}
Physically, the drag parameter $\gamma$ comes from scatterings with both soft and hard momentum exchanges while $\alpha$ is dominated by soft scatterings that rotate the heavy quark color.

\section{Langevin dynamics and decoherence}
\label{sec:Langevin_decoherence}
\subsection{What is $\langle\bm t^a\rangle?$}
Having derived the Langevin equation \eqref{eq:color_Langevin}, let us investigate its property in more detail.
Hereafter, I regard $\langle\vec {\bm p}\rangle$ and $\langle \vec {\bm x}\rangle$ as classical variables $\vec p$ and $\vec x$.
It is not trivial whether to identify $\langle {\bm t^a} \rangle$ as a classical variable $t^a$ satisfying the constraint $t^a t^a =\frac{1}{2}\left(1-\frac{1}{N_{\rm c}}\right)$.
Note that $t^a$ is now a ($N_{\rm c}^2-1$)-dimensional vector and is different from a matrix $\hat t^a$.
Actually, this identification contradicts the equipartition \eqref{eq:Edot} derived from the master equation because it results in
\begin{eqnarray}
\frac{d}{dt}\left\langle\frac{\vec p^2}{2M}\right\rangle_{\xi}
= -\frac{C_{\rm F}\gamma}{MT}\left(
\left\langle\frac{\vec p^2}{2M}\right\rangle_{\xi}
-\frac{3T}{2}(1+\epsilon)\frac{t^at^a}{C_{\rm F}}
\right),
\end{eqnarray}
with $t^at^a=\frac{1}{2}\left(1-\frac{1}{N_{\rm c}}\right)<C_{\rm F}$.
Therefore with the identification $t^a=\langle {\bm t^a}\rangle$ as a classical variable, the heavy quarks cannot get thermalized correctly.
One easy solution, which turns out to be correct, is to impose a constraint $t^at^a=\langle {\bm t^a}{\bm t^a}\rangle=C_{\rm F}\neq\langle {\bm t^a}\rangle\langle{\bm t^a}\rangle$, independently of the identification $t^a=\langle {\bm t^a} \rangle$.
The problem of this quick solution is that the connection to ${\bm \rho}_{\rm color}(t;\zeta)$ is seemingly lost.
I will see how to interpret the constraint $t^at^a=\langle {\bm t^a}{\bm t^a}\rangle=C_{\rm F}$.

To reach a deeper understanding, let us recall the physical meaning of $\langle{\bm t^a}\rangle$.
By definition, $\langle{\bm t^a}\rangle$ is a quantum expectation value of the heavy quark color charge and it couples to the time evolution of {\it classical} or {\it macroscopic} variable $\vec p$.
This indicates that $\vec p$ can be regarded as a macroscopic variable in the so-called {\it Schr\"odinger's cat state}.
To determine the momentum update, the actual value for $\vec f^a(t)\langle {\bm t^a} \rangle$ should be the one that is as if measured.
Otherwise, the noise term $\vec f^a(t)\langle {\bm t^a} \rangle$ only describes the thermal fluctuation and lacks the quantum fluctuation in the color space.

\subsection{``Measurement" of momentum kicks}
In order to take into account the quantum fluctuation, I introduce a notation $\left[\mathcal O \right]_{\rm meas}$ which takes one of the possible values of the observable $\mathcal O$ in a measurement with respect to a pure state ${\bm \rho}_{\rm color}(t;\zeta)$.
Here I examine how $\vec f^a(t){\bm t^a}$ is measured.
First, let us parametrize $\vec f^a(t)=\vec f(t)n^a(t)$ with normalization $n^a(t)n^a(t)=1$.
To be consistent with the statistical property of $\vec f^a$, the independent white noises $\vec f$ and $n^a$ must satisfy $\langle f_i(t) f_j(t')\rangle_T=(N_{\rm c}^2-1)\gamma(1+\epsilon)\delta(t-t')\delta_{ij}$ and $\langle n^a(t)n^b(t)\rangle_T=\frac{1}{N_{\rm c}^2-1}\delta^{ab}$.
Then $\left[\vec f^a(t){\bm t^a}\right]_{\rm meas}$ takes values as
\begin{eqnarray}
\left[\vec f^a(t){\bm t^a}\right]_{\rm meas}
=\vec f(t)\times \left\{\frac{1}{2}, 0, \cdots, 0, -\frac{1}{2} \right\}.
\end{eqnarray}
There are $N_{\rm c}-2$ zero modes for $\vec f^a(t){\bm t^a}$ in the fundamental representation.
The probability that a pure state ${\bm \rho}_{\rm color}(t;\zeta)$ shrinks to each eigenstate depends on the details of both the pure state and the eigenstate.
The quantum fluctuation can also be described using the Wigner function.
See Appendix \ref{sec:Wigner} for a brief summary.

By interpreting the momentum update as
\begin{eqnarray}
\label{eq:pdot_xi_meas}
\frac{d}{dt} \vec p
&=&-\frac{C_{\rm F}\gamma}{2MT}\vec p + \vec f(t)\left[n^a(t)\bm{t^a}\right]_{\rm meas},
\end{eqnarray}
one can take into account the quantum fluctuation in the color space.
In this interpretation, the noise strength due to the quantum fluctuation is evaluated as
\begin{eqnarray}
\vec f(t)^2{\rm Tr}_{\rm color}\left\{
{\bm \rho}_{\rm color}(t;\zeta)
\left(n^a(t)\bm{t^a}\right)^2
\right\}.
\end{eqnarray}
Together with the thermal fluctuation, the noise strength becomes
\begin{eqnarray}
&&\left\langle\vec f(t)^2
{\rm Tr}_{\rm color}\left\{
{\bm \rho}_{\rm color}(t;\zeta)
\left(n^a(t)\bm{t^a}\right)^2
\right\}\right\rangle_{\xi}\nonumber\\
&&= 3\gamma(1+\epsilon)\delta(0)
\llangle\bm{t^a}\bm{t^a}\rrangle
=3C_{\rm F}\gamma(1+\epsilon)\delta(0),
\end{eqnarray}
and the equipartition of heavy quarks is realized correctly.
In an original derivation, $\delta(0)=1/dt$ with $dt$ being a {\it mathematical} discretization time scale.
The discretization time scale $dt$ is shorter than any physical time scales of the heavy quark and is thus sent to $dt\to 0$.
One will see, however, that there is an appropriate discretization time scale $\Delta t$ at which the ``measurement" makes sense.
Therefore, physically, one should take $\delta(0)=1/\Delta t$.

In the above, the factorization of $\vec f^a(t)=\vec f(t)n^a(t)$ is essential for the simultaneous measurements of $f^a_x{\bm t^a}$, $f^a_y{\bm t^a}$, and $f^a_z{\bm t^a}$.
In general, $\vec f^a(t)\equiv -\vec\nabla \xi^a(t,\vec x=\langle \vec {\bm x}\rangle(t))$ does not factorize in such a way.
However, I need to assume the factorization only at the time scale $\Delta t$, during which, as one will see shortly, several soft scatterings take place.
Therefore, the factorization is assumed only for the statistical property of the total momentum kick accumulated during $\Delta t$.

\subsection{Decoherence time scale}

\begin{figure}[t]
\includegraphics[angle=90, clip,width=8.0cm]{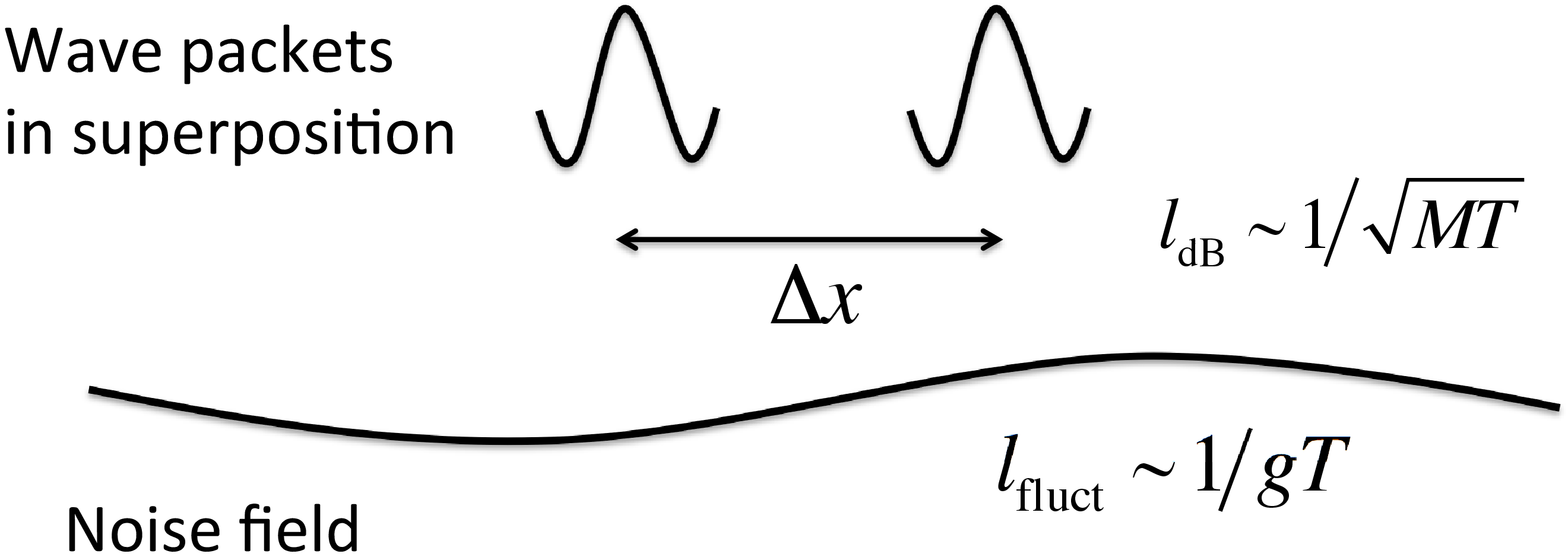}
\label{fig:decoherence}
\caption{Length scales of wave packets in a superposition state and the noise field.}
\end{figure}

Here I show that the quantum fluctuation of $\vec f^a(t){\bm t^a}$ becomes relevant as a consequence of decoherence.
For a moment, let us get back to the quantum master equation for the heavy quark \eqref{eq:master_ctraced}.
The second line of Eq.~\eqref{eq:master_ctraced} describes decoherence of heavy quark wave function.
When the heavy quark wave function is localized compared to the Debye screening length $l_{\rm fluct}\sim 1/gT$, $F_1(\vec x-\vec y)$ can be Taylor expanded and the decoherence term can be approximated as
\begin{eqnarray}
\frac{1}{2}C_{\rm F}\gamma(1+\epsilon)(\vec x-\vec y)^2\bar\rho(t,x,y).
\end{eqnarray}
This term tells us that a wave function of size $\Delta x \ll l_{\rm fluct}$ loses coherence in a time scale $\tau_{\rm dec}\sim 1/C_{\rm F}\gamma (\Delta x)^2$.
If the reduced Planck constant $\hbar$ is recovered, $\tau_{\rm dec}\sim \hbar^2/C_{\rm F}\gamma (\Delta x)^2$ indicating the quantum nature of decoherence. 

Let us apply this decoherence time formula to the case of the decoherence of the macroscopic superposition in the momentum space.
The decoherence time gives an estimate for an appropriate time step $\Delta t\approx \tau_{\rm dec}$ in the classical Langevin dynamics.
Figure \ref{fig:decoherence} depicts hierarchy of the length scales I consider.
When the momentum difference of the wave packets in the macroscopic superposition state is $\Delta p$, it propagates to the distance between them $\Delta x\sim (\Delta p/M)\Delta t$ in time $\Delta t$.
The momentum difference due to the noise is $\Delta p \sim |\vec f(t)| (2N_{\rm c})^{-1/2}\Delta t\sim \sqrt{C_{\rm F}\gamma \Delta t}$.
Therefore I get $\Delta x\sim \left(C_{\rm F}\gamma\right)^{1/2}(\Delta t)^{3/2}/M$ for the distance between the wave packets.
As the time step $\Delta t$ lengthens, the distance between them $\Delta x$ increases and thus the decoherence time $\tau_{\rm dec}$ decreases.
The time step $\Delta t$ of the classical Langevin dynamics should be comparable to the decoherence time scale $\tau_{\rm dec}$ during which the quantum interference between the wave packets is destroyed.
Equating $\Delta t\approx\tau_{\rm dec}\sim 1/C_{\rm F}\gamma (\Delta x)^2$, I obtain
\begin{eqnarray}
\Delta t\sim \sqrt{\frac{M}{C_{\rm F}\gamma}}
\sim \sqrt{N_{\rm c}}\sqrt{\frac{M}{(g^2N_{\rm c})^2 \ln (1/g)T^3}}.
\end{eqnarray}
With explicit dependence on $\hbar$, one can see the quantum origin of the time scale $\Delta t\sim \sqrt{\hbar M/C_{\rm F}\gamma}$. 
In this time scale $\Delta t$, the decoherence turns the quantum superposition in the Schr\"odinger's cat state into a probabilistic mixture of the classical states.

There is an alternative view to the time scale $\Delta t$ from the uncertainty principle \cite{BinWu}.
Here let us recover the reduced Planck constant $\hbar$.
Starting from the same phase space point, {\it the classical trajectories in superposition} are close to each other by $\Delta x\sim (\Delta p/M)\Delta t$ and $\Delta p\sim\sqrt{C_{\rm F}\gamma \Delta t}$ after time $\Delta t$.
Those trajectories get separated by $\Delta x\Delta p\sim (C_{\rm F}\gamma/M)(\Delta t)^2 \sim \hbar$ at $\Delta t\sim \sqrt{\hbar M/C_{\rm F}\gamma}$ and occupy different phase space volume elements.

The appropriate time step $\Delta t$ depends on the mass and the momentum diffusion constant of heavy quark.
\footnote{
One should also compare $\Delta t$ with the hard collision time scale $\tau_{\rm hard}\sim 1/(N_{\rm c}^2-1)g^4\ln (1/g) T\sim 1/(g^2 N_{\rm c})^2T$, which provides a conventional lower limit of time discretization.
Neglecting the logarithms $\ln (1/g)$, $\Delta t > \tau_{\rm hard}$ corresponds to $N_{\rm c}(M/T)(g^2 N_{\rm c})^2 > 1$.
}
This is much shorter than the relaxation time of the heavy quark momentum $\Delta t \ll\tau_{\rm relax}\sim MT/C_{\rm F}\gamma$ and thus is consistent with the above analysis in which the dissipative term with $\vec F_2$ is ignored.
Using this time scale, the distance between the heavy quark wave packets in superposition is $\Delta x\sim (C_{\rm F}\gamma M)^{-1/4}$ so that $\Delta x\ll l_{\rm fluct}$ also holds consistently.
Moreover, the wave function is much more extended than the thermal de Broglie length $\Delta x\gg l_{\rm dB}$, which is a typical size of a wave packet.

\subsection{Color space dynamics}
The appropriate discretization time scale for the classical Langevin dynamics is $\Delta t$.
However, the typical time scale for the color space dynamics is much shorter $\tau_{\rm color}\sim 1/\alpha\sim 1/g^2T$, which is the interval of soft scatterings.
Therefore, in a time scale $\Delta t\gg\tau_{\rm color}$, $\hat\rho_{\rm color}(t)$ is updated, in effect, by a random ${\rm SU}(N_{\rm c})$ matrix.
Replacing $\hat\rho_{\rm color}(t;\zeta)$ by a random pure state, which subsequently shrinks to one of the eigenstates of $n^a(t)\hat t^a$, the heavy quark color degrees of freedom are no more dynamical degrees of freedom.
For a random pure state, $\left[n^a(t)\bm{t^a}\right]_{\rm meas}$ takes values $\left\{1/2,0,\cdots,0,-1/2\right\}$ with equal probabilities.
This allows us to treat $\left[n^a(t)\bm{t^a}\right]_{\rm meas}$ as a stochastic variable with variance $1/2N_{\rm c}$. 
In this way, the normal Langevin equation without color degrees of freedom is derived at time scale of $\Delta t\sim \sqrt{M/C_{\rm F}\gamma}$:
\begin{eqnarray}
\label{eq:Langevin_nocolor}
\left\{
\begin{aligned}
&\Delta \vec x = \frac{\vec p}{M}\Delta t, \ \ \ 
\Delta \vec p = -\frac{C_{\rm F}\gamma}{2MT}\vec p\Delta t + \vec \eta(t)\Delta t,\\
&\langle \eta_i(t)\eta_j(t)\rangle_T = C_{\rm F}\gamma(1+\epsilon)\delta_{ij}/\Delta t.
\end{aligned}
\right.
\end{eqnarray}

\section{Summary}
\label{sec:summary}
In this paper, I studied the Langevin dynamics of heavy quarks with internal color degrees of freedom.
The Langevin dynamics is not obtained in a closed form in the phase space but as a coupled dynamics in the phase space and in the color space.
The coupling causes the macroscopic superposition state, or the so-called  Schr\"odinger's cat state, in the updates of heavy quark momentum.
The classical variable $\vec p$, which corresponds to the cat, couples to the quantum state in the color space.
The quantum interference in the superposition state is destroyed in a time scale $\Delta t\sim \sqrt{M/C_{\rm F}\gamma}$ by the decoherence and the momentum update should be regarded as one of the possible outcome by a ``measurement''.
Note that the macroscopic superposition and the decoherence in the Langevin dynamics are unique to the non-Abelian interaction.
In this way, I am naturally led to take $\Delta t\sim \sqrt{M/C_{\rm F}\gamma}$ for a discretization time scale of the Langevin dynamics.
It is interesting because the physical quantities of the classical Langevin equation ($M$ and $\gamma$) determine the discretization time scale $\Delta t$ for solving it.
At this time scale $\Delta t$, the color degrees of freedom are expected to be randomized.

There have been extensive phenomenological studies on the heavy quark energy loss and the drag force parameter $C_{\rm F}\gamma \sim 1 T^3$ has been extracted from the experimental data \cite{Rapp:2009my}.
The decoherence of a charm (bottom) quark in the QGP with $T\sim 200 \ {\rm MeV}$ proceeds in a time scale $\Delta t \sim 3 \ {\rm fm} \ (5 \ {\rm fm})$.
These values are not very small compared to the typical lifetime of the QGP in the heavy-ion collisions, $\tau_{\rm QGP}\sim 10 \ {\rm fm}$.
Moreover, it is longer than the time scale of hadronization, which I roughly estimate to be $1/\Lambda_{\rm QCD}\sim 1 \ {\rm fm}$.
This indicates that not only might the classical Langevin equation be inappropriate to describe heavy quarks in the fireball, but also the freezeout process might be able to resolve a heavy quark wave function in the macroscopic superposition state.
Such a wave function is extended about $\Delta x\sim (C_{\rm F}\gamma M)^{-1/4}\sim 0.6 \ {\rm fm} \ (0.45 \ {\rm fm})$ for a charm (bottom) quark and may enhance the probability of recombining heavy quark$\-$antiquark pairs into heavy quarkonia.

To establish a better (semi)classical description for heavy quarks in the heavy-ion collisions, it would be desirable to derive a novel kinetic theory for heavy quarks in which (i) the time scale below $\Delta t\sim\sqrt{M/C_{\rm F}\gamma}$ can be resolved and (ii) the superposition of wave packets and their decoherence are effectively described.
The Kadanoff-Baym equation for 4-dimensional Wigner function can be helpful for this project \cite{KadanoffBaym}.

Finally, let us remark on the similarity between the discretization time scale $\Delta t\sim \sqrt{M/C_{\rm F}\gamma}$ and the formation time of induced gluon radiation off an energetic parton $\tau_{\rm form}\sim \omega/k_{\perp}^2\sim \sqrt{\omega/\hat q}$ \cite{Baier:2000mf}.
Here $\omega$ and $k_{\perp}$ are gluon energy and transverse momentum, and $\hat q$ is a transport coefficient characterizing transverse momentum kicks which the energetic parton experiences in the medium.
In both cases, decoherence determines the time scales.

For future prospect, it would be an interesting challenge to simulate the stochastic master equation \eqref{eq:stochastic_master}, possibly in a simplified version without $\vec F_2$ terms.
Without the $\vec F_2$ terms, the stochastic master equation is equivalent to the stochastic Schr\"odinger equation \cite{Akamatsu:2011se}.
The natural time step of the stochastic master equation is $\tau_{\rm color}$ while the classical Langevin dynamics is expected to emerge at longer time scale $\Delta t\gg\tau_{\rm color}$.
Starting from a localized wave packet state, one may ask the following key questions:
\begin{enumerate}
\item Does the thermal fluctuation with color $\xi^a$ generate a superposition state of wave packets?
\item How is the interference between the wave packets destroyed by the decoherence?
\item How does $\hat\rho_{\rm Q}(t,\vec x,\vec y;\xi)$ acquire the effect of decoherence in the corresponding classical phase space distribution?
\end{enumerate}
The answers to these questions will help bridge a gap between the quantum and classical descriptions of heavy quarks in the high-temperature quark-gluon matter.

\section*{Acknowledgements}
I thank Jean-Paul Blaizot, Tetsufumi Hirano, Chiho Nonaka, and Derek Teaney for fruitful discussions in an early stage of this study.
I also thank Bin Wu for suggesting an alternative interpretation of the decoherence time scale, during the molecule-type workshop ``Selected Topics in the Physics of the Quark-Gluon Plasma and Ultrarelativistic Heavy Ion Collisions," at Yukawa Institute for Theoretical Physics, Kyoto University.

\appendix
\section{Thermal fluctuation in the influence functional}
\label{sec:thermal_fluctuation}
In the Markov limit, the influence functional can be expanded as $S_{\rm IF}=S_{\rm pot}+S_{\rm fluct}+S_{\rm diss}+\cdots$ \cite{Akamatsu:2014qsa}.
The influence functional gives a part of path-integral weight $e^{iS_{\rm IF}}$ for the propagation of the reduced density matrix.
Explicit forms of $S_{\rm fluct}$ and $S_{\rm diss}$ are
\begin{eqnarray}
iS_{\rm fluct}&=&-\frac{1}{2}\int_{t_0} dt\int d^3xd^3yD(\vec x-\vec y)\nonumber\\
&& \ \times\left(
\rho^{a}_1,\ \rho^{a}_2
\right)_{(t,\vec x)}
\left[
\begin{array}{cc}
-1 & 1 \\
1 & -1
\end{array}
\right]
\left(
\begin{array}{c}
\rho^{a}_1\\
\rho^{a}_2
\end{array}
\right)_{(t,\vec y)},\\
iS_{\rm diss}&=&\frac{i}{4T}\int_{t_0} dt\int d^3xd^3y\vec\nabla_xD(\vec x-\vec y)\nonumber\\
&& \ \cdot\left(
\rho^{a}_1,\ \rho^{a}_2
\right)_{(t,\vec x)}
\left[
\begin{array}{cc}
-1 & -1 \\
1  &  1
\end{array}
\right]
\left(
\begin{array}{c}
\vec j^{a}_1\\
\vec j^{a}_2
\end{array}
\right)_{(t,\vec y)}.
\end{eqnarray}
Here $(\rho^a, \vec j^a)$ is nonrelativistic heavy quark color current and labels 1 and 2 denote the Schwinger-Keldysh contour.
Integrating by parts, the spatial derivatives in $\vec j^a$, $S_{\rm diss}$ contains terms of similar structure to $S_{\rm fluct}$.
Let us call this contribution $S_{\rm diss}'$.
Then
\begin{eqnarray}
iS_{\rm fluct}+iS_{\rm diss}'&=&\frac{1}{2}\int_{t_0} dt\int d^3xd^3yF_1(\vec x-\vec y)\\
&& \ \times\left(
\rho^{a}_1,\ \rho^{a}_2
\right)_{(t,\vec x)}
\left[
\begin{array}{cc}
-1 & 1 \\
1 & -1
\end{array}
\right]
\left(
\begin{array}{c}
\rho^{a}_1\\
\rho^{a}_2
\end{array}
\right)_{(t,\vec y)}.\nonumber
\end{eqnarray}
Since $F_1(\vec r)$ is positive definite for $M\gg T$, the thermal fluctuation and a part of the dissipation $e^{iS_{\rm fluct}+iS_{\rm diss}'}$ can be expressed using a Gaussian white noise $\xi^a$ as
\begin{eqnarray}
&&e^{iS_{\rm fluct}+iS_{\rm diss}'}\\
&&=\left\langle
\exp\left[-i\int_{t^0}dt\int d^3x \xi^a(t,\vec x)\left(\rho^a_1(t,\vec x)-\rho^a_2(t,\vec x)\right)\right]
\right\rangle_{\xi}, \nonumber \\
&& \langle \xi^a(t,\vec x)\xi^b(s,\vec y)\rangle_T=F_1(\vec x-\vec y)\delta(t-s)\delta^{ab}.
\end{eqnarray}
This gives nothing but the stochastic propagation in Eq.~\eqref{eq:master_xi}.

\section{Wigner function and quantum fluctuation}
\label{sec:Wigner}
I investigated how to reconcile the kinetic equilibration of heavy quarks with the Langevin equation \eqref{eq:color_Langevin}.
I found that the quantum fluctuation in the color space is essential and proposed an interpretation \eqref{eq:pdot_xi_meas}.

Using the Wigner function, I can also describe the same physics and discuss the effect of quantum fluctuation.
The stochastic master equation \eqref{eq:stochastic_master} with explicit $\hbar$ is obtained by substituting $(\vec\nabla_x^2-\vec\nabla_y^2)/2M\to\hbar(\vec\nabla_x^2-\vec\nabla_y^2)/2M$, $F_1\to F_1/\hbar^2$, and $\xi^a\to \xi^a/\hbar$.
The Wigner function is defined by
\begin{eqnarray}
\hat W_{\rm Q}(t,\vec r,\vec p;\xi)\equiv
\int d^3s e^{-i\frac{\vec p\cdot\vec s}{\hbar}}
\hat\rho_{\rm Q}\left(t,\vec r+\frac{\vec s}{2}, \vec r-\frac{\vec s}{2};\xi\right).
\end{eqnarray}
Using $\hat W_{\rm Q}$, the color-averaged Wigner function is given by $\overline W_{\rm Q}={\rm Tr}_{\rm color}\left\{\hat W_{\rm Q}\right\}$.
Taking the limit $\hbar\to 0$ in the master equation \eqref{eq:stochastic_master}, I obtain a Kramers equation for $\overline W_{\rm Q}$:
\begin{eqnarray}
&&\left(\frac{\partial}{\partial t}
 + \frac{\vec p}{M}\cdot \vec\nabla_r\right)
\overline W_{\rm Q}(t,\vec r,\vec p;\xi)\nonumber\\
&&=\frac{C_{\rm F}\gamma}{2MT}\frac{\partial}{\partial\vec p}\cdot
\left(\vec p + MT(1+\epsilon)\frac{\partial}{\partial\vec p}\right)
\overline W_{\rm Q}(t,\vec r,\vec p;\xi)\nonumber\\
&& \ \ \ -\vec f^a(t,\vec r)\cdot\frac{\partial}{\partial\vec p}
{\rm Tr}_{\rm color}\left\{\hat W_{\rm Q}(t,\vec r,\vec p;\xi)\hat t^a\right\},
\end{eqnarray}
with $\vec f^a(t,\vec r)\equiv -\vec\nabla \xi^a(t,\vec r)$.
The Kramers equation is not obtained as a closed form of $\overline W_{\rm Q}$ but rather depends on the color state of $\hat W_{\rm Q}$.
Let us make the equivalent assumptions which have been made previously: 
\begin{enumerate}
\item $\overline W_{\rm Q}$ is localized in the phase space.
This allows us to replace $\vec f^a(t,\vec r)$ by $\vec f^a(t)$.
\item $\hat W_{\rm Q}$ can be factorized into the Wigner function in the phase space and that in the color space: $\hat W_{\rm Q}(t,\vec r,\vec p;\xi)=\overline W_{\rm Q}(t,\vec r,\vec p;\xi)\cdot \hat W_{\rm Q}^{\rm color}(t;\xi)$.
\item The momentum kick at the position of the heavy quark can also be factorized: $\vec f^a(t)=\vec f(t)n^a(t)$.
\end{enumerate}
In order to take the quantum fluctuation into account, the last term of the Kramers equation can be understood as
\begin{eqnarray}
&&-\vec f(t)\left[n^a(t)\hat t^a\right]_{\rm meas}
\cdot\frac{\partial}{\partial\vec p}\overline W_{\rm Q}(t,\vec r,\vec p;\xi),
\end{eqnarray}
where $\left[n^a(t)\hat t^a\right]_{\rm meas}$ is evaluated by the color state of $\hat W_{\rm Q}^{\rm color}(t;\xi)$.
The dynamics of $\hat W_{\rm Q}^{\rm color}(t;\xi)$ is the same with that of $\hat \rho_{\rm color}(t;\zeta)$ in Eq.~\eqref{eq:colormaster_xi}.

\bibliographystyle{apsrev}

\end{document}